\documentclass[12pt]{iopart}
\usepackage{iopams}
\usepackage{graphicx}
\begin{document}

\title[Quantum transport in presence of a background]{Ballistic electron
quantum transport in presence of a disordered background}

\author{Valentin V. Sokolov}

\address{Budker Institute of Nuclear Physics 630090
Novosibirsk, acad. Lavrentiev prospect 11, Russia; Novosibirsk
State Technical University}
\ead{V.V.Sokolov@inp.nsk.su}

\begin{abstract}
Effect of a complicated many-body environment is analyzed  on
the electron random scattering by a 2D mesoscopic open ballistic
structure. A new mechanism of decoherence is proposed. The temperature
of the environment is supposed to be zero whereas the energy of the
incoming particle $E_{in}$ can be close to or somewhat
above the Fermi surface in the environment. The single-particle doorway
resonance states excited in the structure via  external channels are
damped not only because of escape through such channels but also due
to the ulterior population of the long-lived environmental
states. Transmission of an electron with a given incoming $E_{in}$ through the
structure turns out to be an incoherent sum of the flow formed
by the interfering damped doorway resonances and the retarded
flow of the particles re-emitted into the structure by the
environment. Though the number of the particles is
conserved in each individual event of transmission, there exists
a probability that some part of the electron's energy can
be absorbed due to environmental many-body effects. In such a
case the electron can disappear from the resonance energy
interval and elude observation at the fixed transmission energy
$E_{in}$ thus resulting in seeming loss of particles, violation of the
time reversal symmetry and, as a consequence, suppression of the
weak localization. The both decoherence and absorption
phenomena are treated within the framework of a unit microscopic
model based on the general theory of the resonance scattering.
All the effects discussed are controlled by the only parameter:
the spreading width of the doorway resonances, that uniquely
determines the decoherence rate.
\end{abstract}

\submitto{\JPA}
\maketitle

\section{Introduction}
With the advent of the ability to fabricate mesoscopic analogs
of the classical billiards - eminent systems often used to
illustrate the characteristic features of the classical
dynamical chaos, the opportunity appeared to directly observe
signatures of chaos in classically chaotic quantum systems. An
excellent possibility thus has arisen to verify experimentally
the theoretical concepts developed in numerous theoretical
investigations of "quantum chaos" phenomena. Extensive study of
the electron transport through ballistic meso-structures
\cite{Jalabert1990,Marcus1992} (see also
\cite{Beenakker1997,Alhassid2000} and references therein) have
fully confirmed correctness of the basic ideas
\cite{Izrailev1990,Gutzwiller1991,Haake1991} of the theory of
the chaotic quantum interference as well as relevance
\cite{Mello1994,Beenakker1997,Alhassid2000} of the
random-matrix approach
\cite{Verbaarschot1985,Mello1985,Mehta1991,Fyodorov1997} to the problem of
the universal fluctuations in open mesoscopic set-ups.
Nevertheless, experiments with the ballistic quantum dots
\cite{Marcus1993, Bird1995,Clarke1995,Huibers1998} reveal
noticeable and persisting up to zero temperature loss of the
quantum-mechanical coherence in contravention of predictions of
the standard semiclassical and random-matrix scattering
theories.

A number of different methods of accounting for the decoherence
in the ballistic quantum transport processes have been
suggested. In the phenomenological voltage-probe model that
goes back to the B\"uttiker's papers \cite{Buettiker1986} a
subsidiary dephasing lead with vanishing mean current is
attached to the cavity. Generally speaking, the fictitious lead
can support an arbitrary number $M_{\phi}$ of channels with
some transmission coefficients $T_{\phi}$. To get rid of the
arising ambiguity a special procedure has been worked out in
\cite{Brouwer1997ii}: the limit
$M_{\phi}\!\rightarrow\!\infty,\, T_{\phi}\!\rightarrow 0$ has
been considered where the product of these two numbers remains
constant and is fixed via the connection
$\gamma_{\phi}=\frac{2\pi}{D}\,M_{\phi}\!T_{\phi}$ by the
decoherence rate $\gamma_{\phi}$ - the only parameter of the
actual physical meaning. (Here $D$ stands for the mean level
spacing in the mesoscopic cavity.) It should be noted in this
connection that the construction proposed implicitly suggests a
complicated internal structure of the dephasing probe which
should, in particular, possess a dense energy spectrum with the
mean level spacing $d\ll D$. Otherwise, the assumed limit could
hardly be physically justified. If so, the typical time
$\tau_d=\frac{2\pi}{d}$ spent by the electron inside the probe
constitutes a new time scale different from the mean time delay
$\tau_D=\frac{2\pi}{D}$ in the cavity.

Another approach has been proposed in \cite{Efetov1995} where
the decoherence phenomenon is linked to the electron absorption.
The absorption can readily be modelled by
including in the Hamiltonian a spatially uniform imaginary
potential whose strength $-\frac{i}{2}\gamma_{\phi}$ is
directly connected to the decoherence rate. Obviously, the
model in which the number of particles is not conserved
violates unitarity of the scattering matrix. Nevertheless, the
authors of the ref. \cite{Brouwer1997ii} have managed to accord
and combine the two mentioned models by compulsory restoring
{\it in average} conservation of the number of particles with
a given energy so that the only effect of such a
dephasing probe consists in erasing the phase memory. However,
in a wider aspect, a voltage probe allows for energy dissipation
also. An elucidative comparison of properties of the dissipative
voltage probes on the one hand and the energy conserving dephasing
probes on the other hand has been given in \cite{Buettiker2006}.
Still, the probe models formulated in such a way are not, as has been
noticed in \cite{Beenakker2005}, entirely satisfactory.
Indeed, only the mean value of the total electron flow through the
probe can be forced to vanish. The unitarity of the
scattering matrix is not perfectly restored and the number of
electrons is not conserved in each individual act of
scattering.

An alternative phenomenological model of the dephasing has
been therefore proposed in \cite{Beenakker2005} with a closed
long dephasing stub instead of an opening lead.
Thereby unitarity of the scattering matrix is guaranteed and
none of the electrons is lost at any individual measurement. At
last, the dephasing in the stub has been supposed to be
induced by a spatially random time-dependent external electric
field that breaks the phase coherence in the way similar to
that known from the theory of dephasing in bulk disordered
conductors \cite{Altshuler1985}. As a result the electrons once
penetrated the stub return back in the cavity without any phase
memory.

In spite of the advantages of the stub model the
necessity of introducing {\it ad hoc} an external time dependent
potential seems to be somewhat artificial. In this paper we discuss
within the framework of a unit microscopic model a
simple alternative mechanism of the decoherence and dissipation
phenomena induced by a time-independent weak interaction with a
disordered environment. The disorder arises due to relatively 
rare irregular impurities in the semicondactor heterostructure to 
whose interface region the electrons are confined. We suggest that though
the electron's mean free path exceeds the size of the dot there exists
some finite probability for an electron to be scattered by such an impurity
during its stay inside the dot. Still such a scattering cannot by itself
destroy the phase coherence. Our model takes into account also unavoidable
energy averaging because of the finite accuracy with
which electron's energy can be measured in any individual scattering
event. In Sec. II. we present our
model and describe the fine-scale fragmentation of the electron
resonant states in a mesocsopic structure induced by
interaction with a disordered environment. A resonant
representation of the scattering matrix in the presence of a
disordered environment is derived. In Sec. III. we carry out
the fine-structure averaging and show how this averaging
results in the decoherence. The cases
of isolated and overlapping resonances are analysed at zero
temperature of environment. The role of environmental many-body
effects and the energy absorption are then considered
when the energy of incoming electron $E_{in}$ noticeably exceeds the Fermi energy
in the environment. Ensemble averaging and suppression of the weak
localization are discussed in Sec.IV. We summarize our findings
in Sec. V.

\section{Fragmentation of the doorway resonance states due to interaction
with a disordered background}
Let $H^{(s)}$ be the Hamiltonian of an ideal ballistic mesocsopic
cavity with perfectly reflecting walls and no environment. Let us
suppose that the cavity is attached to two long leads which
support altogether $M$ transversal (channel) modes. An open
system of such a kind is described by the effective
non-Hermitian Hamiltonian ${\cal H}^{(s)}= H^{(s)}-\frac{i}{2}A
A^{\dag}$ \cite{Verbaarschot1985,Sokolov1989,Sokolov1992} where
the rectangular matrix $A$ consists of $M^{(s)}$ column vectors
of transition amplitudes between $N^{(s)}$ internal states excited
in the cavity during the scattering at some electron energy $E$ and
$M^{(s)}$ channel states. Let us further suppose that our cavity
is imbedded in a many-body environment described by the
$N^{(e)}\times N^{(e)}$ {\it Hermitian} Hamiltonian matrix $H^{(e)}$
with a very small mean level spacing $\delta$, to which our system
is coupled via a rectangular matrix $V$. The dimension $N^{(e)}\ggg N^{(s)}$
so that the spacing $\delta$ fixes the smallest energy scale. This spacing
will be kept finite throughout the paper to ensure unitarity of the
scattering matrix $S(E)=I-i{\cal T}(E)$. The total system: the open
cavity interacting with the environment, is described by the extended
non-Hermitian effective Hamiltonian
\begin{equation}\label{ExtHam}
{\cal H}=\left(
\begin{array}{cc}
{\cal H}^{(s)} & V^{\dag} \\
V & H^{(e)}\\
\end{array}%
\right).
\end{equation}
The corresponding transition matrix equals ${\cal T}(E)=
A^{\dag}{\cal G}_D(E)A$
where ${\cal G}_D(E)$ stands for the upper left block
\begin{equation}\label{DProp}
{\cal G}_D(E)=\frac{I}{E-{\cal H}^{(s)}-\Sigma(E)}
\end{equation}
of the resolvent ${\cal G}(E)=\frac{I}{E-{\cal H}}$ of the
extended non-Hermitian Hamiltonian ${\cal H}$. The subscript
$D$ means "doorway" and marks the resonance states in the ideal
open cavity, that are directly connected to the scattering
channels \cite{Feshbach1968,Sokolov1992}. In zero approximation
$V\equiv 0$, only these states are unstable and have complex
eigenenergies, ${\cal E}_n=\varepsilon_n-\frac{i}{2}\Gamma_n$.
The environment states get excess to the leads only due to the
mixing with the doorway resonances exclusively through which they
can be excited or relax. The matrix
$\Sigma(E)=V^{\dag}\frac{1}{E-H^{(e)}}V$ accounts
for transitions cavity $\leftrightarrow$ environment and
remains Hermitian (and, correspondingly, the scattering matrix
remains unitary) as long as the energy spectrum of the
environment is discrete, i.e. the mean level spacing
$\delta\neq 0$.

In the mean-field single-particle approximation $H^{(e)}
\approx H_{sp}^{(e)}$, an electron penetrating into the environment
moves in a mean field that is random because of impurities.
So we suppose that the coupling matrix elements are random Gaussian
quantities,
\begin{equation}\label{Rand_V}
\langle V_{\mu m}\rangle=0\,,\,\,\langle V_{\mu m}^*V_{\nu n}\rangle=
\frac{1}{2}\Gamma_s\frac{d}{\pi}\delta_{\mu\nu}\delta_{m n}\,.
\end{equation}
The subscripts $m, n$ and $\mu, \nu$ mark the doorway and the
background single-particle states respectively. As the
condition $\delta\neq 0$ holds the quasi-particle's energy spectrum
is discrete with a mean level spacing $d$
that satisfies the inequalities $\delta\lll d\ll D$. The spreading width $\Gamma_s=2\pi\frac{\langle |V|^2\rangle}{d}$
characterizes the fine-scale fragmentation of the
doorway states because of the coupling to the environment. It is
understood that the spreading width $\Gamma_s\gg d$.

Using eq. (\ref{Rand_V}) we can substitute with the accuracy $1/N_{sp}^{(e)}$ the matrix $\Sigma(E)$ in (\ref{DProp})  by the
averaged value:
\begin{equation}\label{EvSigma}
\Sigma(E)\Rightarrow\frac{1}{2}\Gamma_s\,g(E);\quad g(E)=
\frac{d}{\pi}Tr\frac{1}{E-H_{sp}^{(e)}}\,.
\end{equation}
Here $N_{sp}^{(e)}$ is the dimension of the Hilbert space of a
quasi-electron in the environment. Since the spectrum of the quasi-particle's Hamiltonian $H_{sp}^{(e)}$ is discrete the loop
function $g(E)$ is real so that the V-averaging does not destroy unitarity
of the scattering matrix $S(E)$.

The transition amplitudes reduce after that to a sum of the doorway
resonant contributions
\begin{equation}\label{T_DwayRep}
{\cal T}^{ab}(E)=\sum_n\frac{{\cal A}_n^a {\cal A}_n^b}{E-{\cal E}_n
-\frac{1}{2}\Gamma_s g(E)}\equiv
\sum_n\frac{{\cal A}_n^a {\cal A}_n^b}{{\cal D}_n(E)}\,.
\end{equation}
For some time, we restrict ourselves to the case of the systems
with time reversal symmetry. The decay amplitudes $A_n^a$ are
real in this case and the matrix of the non-Hermitian effective
Hamiltonian ${\cal H}^{(s)}$ is symmetric. The  amplitudes
${\cal A}_n^a$ in Eq. (\ref{T_DwayRep}) are the matrix elements
of the coupling matrix ${\cal A}=\Psi^T A$ with $\Psi$ being
the complex orthogonal ($\Psi^T\Psi=1$) matrix of the
eigenstates of the effective Hamiltonian ${\cal H}^{(s)}$.
Therefore, unlike the real elements of the matrix $A$, those of
the matrix ${\cal A}$ are complex quantities
\cite{Sokolov1989,Sokolov1992}.

The exact resonance spectrum $\{{\cal E}_{\alpha}\}$ is now
found by solving $N^{(s)}$ independent equations
\begin{equation}\label{ExRes}
{{\cal D}_n({\cal E}^n_{\nu})}=
{\cal E}^n_{\nu}-{\cal E}_n-\frac{1}{2}\Gamma_s g({\cal E}^n_{\nu})=0.
\end{equation}
Each doorway state is thus fragmented onto $\sim\Gamma_s/d\gg
1$ narrow fine-scale resonances. Finally, transition amplitudes
can be represented as coherent sums of
$N_{sp}^{(e)}=N^{(s)}\cdot\Gamma_s/d $ interfering resonant
contributions
\begin{equation}\label{ExactResRap}
{\cal T}^{ab}(E)=
\sum_{\alpha}\frac{{\cal A}_{\alpha}^a
{\cal A}_{\alpha}^b}{E-{\cal E}_{\alpha}}\,.
\end{equation}
As distinct from the case of the ideal cavity the interference
pattern depends now on two additional parameters: the spreading
width $\Gamma_s$ and the fine-scale level spacing $d$. Up to this
point our consideration has been general enough and is applicable
in quite a wide scope. In what follows we reexamine in this framework
the problems of the decoherence and absorption in 2D mesoscopic
devices.

\section{Energy averaging over the fine-structure scale}
Let us now take into account that the energy resolution $\Delta E$ is 
not perfect and does not allow for resolving the fine structure of
the doorway resonances, $d\ll\Delta E$ though of course $\Delta
E\ll D$. Then only averaged cross sections
\begin{equation}\label{AvCrossSec}
\overline{\sigma^{ab}(E)}=\frac{1}{\Delta E}\int_{E-
\frac{1}{2}\Delta E}^{E+\frac{1}{2}\Delta E}d E'\,\sigma^{ab}(E')
\end{equation}
are observed. To carry out the energy averaging explicitly, we
neglect the level fluctuations on the fine-structure scale and
assume the uniform spectrum, $\varepsilon_{\mu}=\mu d$ ({\it
the picket fence approximation}). This yields immediately
$g(E)=\cot\left(\frac{\pi E}{d}\right)\,$.

\subsection{Isolated doorway resonance}
In the case of an isolated doorway resonance with the width $\Gamma=
\sum_c\Gamma^c\ll D$, that is situated close to the Fermi energy, $E_{res}=
0$, the transition cross section equals
\begin{equation}\label{TrCrossSec}
\sigma^{ab}(E)=\Big|{\cal T}^{ab}(E)\Big|^2=
\frac{\Gamma^a\Gamma^b}{\left[E-
\frac{1}{2}\Gamma_s\cot\left(\frac{\pi E}{d}\right)\right]^2+
\frac{1}{4}\Gamma^2}
\end{equation}
and the fine-scale energy averaging yields
\begin{equation}\label{AvTrCrossSec}
\overline{\sigma^{ab}(E)}=\frac{\Gamma^a\Gamma^b}{E^2+
\frac{1}{4}\left(\Gamma+\Gamma_s\right)^2}+
\frac{\Gamma^a\Gamma^b}{\Gamma}\frac{\Gamma_s}{E^2+
\frac{1}{4}\left(\Gamma+\Gamma_s\right)^2}\,.
\end{equation}
The phase coherence is destroyed by the averaging and the
result consists of two incoherent contributions. The first one
corresponds to excitation and subsequent decay of the doorway
resonance widened because of leaking into the environment. This
effect is described by shifting in the upper part of the
complex energy plane by the distance $\frac{1}{2}\Gamma_s$. The
second term accounts for the particles re-injected from the
background. There is no net loss of the particles. The
environment looks from outside as a black box which swallows
particles and spits them back in the cavity after some time.

The transport through the cavity is characterized by the
quantity \cite{Beenakker1997,Alhassid2000}
\begin{equation}\label{Transp1}
G(E)=\sum_{a\in 1,d\in 2}\overline{\sigma^{ab}(E)}=
\frac{\Gamma_1\Gamma_2}{\Lambda(E)}+
\frac{\Gamma_1\Gamma_2}{\Gamma_1+\Gamma_2}
\frac{\Gamma_s}{\Lambda(E)}=
T_{12}+\frac{T_{1s}T_{s2}}{T_{1s}+T_{s2}}
\end{equation}
where
$\Lambda(E)=E^2+\frac{1}{4}\left(\Gamma+\Gamma_s\right)^2$. The
second term in the final expression is expressed in terms of
the subsidiary transitions probabilities
\begin{equation}\label{SubTrProb1}
T_{sk}(E)=\frac{\Gamma_s\,\Gamma_k}{\Lambda(E)}\,,\quad
\Gamma_k=\sum_{c\in k}\Gamma^c, \quad k=1, 2,
\quad \Gamma_1+\Gamma_2=\Gamma\,.
\end{equation}
One can interpret Eqs.
(\ref{Transp1}, \ref{SubTrProb1}) by introducing an additional fictitious
$(M^{(s)}+1)$th channel with the transition amplitude
$A^{(s)}=\sqrt{\Gamma_s}$ that connects the resonance state to
the environment. It is easy to check that the fictitious
scattering matrix ${\tilde S}(E)=I-i{\tilde {\cal T}}(E)$ built
in such a way is unitary. The expression (\ref{Transp1}) is formally
identical to that obtained with the B\"uttiker's voltage probe
model \cite{Buettiker1986} of the decoherence phenomenon. At that
the decoherence rate $\gamma_{\phi}=\gamma_s=\frac{2\pi}{D}\Gamma_s=\Gamma_s\tau_D$ is unambiguously defined by the only parameter - the spreading
width of the doorway resonance.

The single-particle approximation used up to now is well
justified only when the scattering energy $E$ is very close to
the Fermi surface in the environment. For higher scattering
energies, many-body effects should be taken into account. They
show up, in particular, in a finite lifetime of the quasi-particle
with the energy $E>E_F=0$. The simplest way to account for this
effect is to attribute some imaginary part to the quasi-particle's
energy, $\varepsilon_{\mu}=\mu d - \frac{i}{2}\Gamma_e$.

The resonant denominator equals then \cite{Sokolov1997}
\begin{equation}\label{AbsProp}
\fl {\cal D}_{res}(E)=E-E_{res}
-\frac{1}{2}\Gamma_s(1-\xi^2)\frac{\eta}{1+\xi^2\eta^2}+
\frac{i}{2}\left(\Gamma+\Gamma_s\xi\frac{1+\eta^2}{1+
\xi^2\eta^2}\right)
\end{equation}
where $E_{res}$ is the position of the doorway resonance and
the following notations are used:
$$\xi=\tanh\left(\frac{\pi\Gamma_e}{2 d}\right),
\quad \eta=\cot\left(\frac{\pi E}{d}\right)\,.$$

The averaged transport cross section $G(E)$ retains still its
form (\ref{Transp1}) but the subsidiary transition
probabilities looks as
\begin{equation}\label{SubTrProb1Abs}
T_{sk}(E)\Rightarrow T_{sk}(E;\kappa)=
\frac{\Gamma_s\,\Gamma_k}{\Lambda(E;\kappa)}
\end{equation}
instead of (\ref{SubTrProb1}). The factor
\begin{equation}\label{eq4}
\frac{1}{\Lambda(E;\kappa)}=\frac{1}{\Lambda(E)}\,\frac{1}{1+
\kappa\frac{\Lambda(E)}{\Gamma\Gamma_s}}
\end{equation}
depends on the new parameter $\kappa$ which accounts for
inelastic effects in the background,
\begin{equation}\label{kappa}
\begin{array}{c}
\kappa=\frac{4\xi}{(1-\xi)^2}=e^{\gamma_e}-1\approx\\
\left\{\begin{array}{l}
\gamma_e\ll 1,\,\,\,\,\,\,\,\textrm{if}\,\,\,\,\,\,\,\,\,\,{\tau}_e\gg\tau_d\,,\\
e^{\gamma_e}\gg 1,\,\,\,\,\, \textrm{if}\,\,\,\,\,\,\,\,\,\,\tau_e<\tau_d\,,\\
\end{array}\right.
\quad(\gamma_e=\tau_d\Gamma_e)\,
\end{array}
\end{equation}
where ${\tau}_e=1/\Gamma_e$ is the lifetime of the quasi-electron
in the environment.

Near the doorway resonance energy $E_{res}$ the influence
of the absorption is negligible within the range $0\leqslant\kappa\lesssim\kappa_c=
\frac{4\Gamma\Gamma_s}{(\Gamma+\Gamma_s)^2}$. The critical
value $\kappa_c$ reaches it's maximum possible, $\kappa_c=1$,
when $\Gamma=\Gamma_s$ and becomes small if one out of the two
widths noticeably exceeds another. In these cases the interval of weak
absorption is very restricted and the absorption begins to play an important
role. If the resonance is so narrow that $\Gamma\ll\Gamma_s$ then $\kappa_c\approx 4\frac{\Gamma}{\Gamma_s}\ll 1$ and the subsidiary probabilities
(\ref{SubTrProb1Abs}) at the resonance energy $E=E_{res}$ and $\kappa\gtrsim
\kappa_c$ are small, $T_{sk}(E=E_{res};\kappa)\approx\frac{16}{\kappa}
\frac{\Gamma\Gamma_k}{\Gamma_s^2}\lesssim\frac{16}{\kappa_c}
\frac{\Gamma\Gamma_k}{\Gamma_s^2}\approx 4\frac{\Gamma_k}{\Gamma_s}\ll 1$.
On the other hand, the very quasi-particle concept is self-consistent only
if $\gamma_e=\tau_d\Gamma_e\lesssim 1$ so that the physically feasible
interval of the strong absorption regime is $\kappa_c\approx 4\frac{\Gamma}{\Gamma_s}\lesssim\kappa\lesssim 1$. In this interval only
the contribution $T_{1 2}(E)$  remains in Eq. (\ref{Transp1}) and our
approach reproduces the result of the Efetov's model
\cite{Efetov1995} with the strength of the imaginary potential
$-\frac{i}{2}\gamma_s$. Under the opposite condition $\Gamma_s\ll\Gamma$
which complies with the assumption that the interaction with the
background is weak the regime of strong absorption arises when $\kappa_c\approx 4\frac{\Gamma_s}{\Gamma}\lesssim\kappa\lesssim 1$. Similar consideration leads
to the Efetov's model again. The decoherence rate equals $\gamma_{\phi}=\gamma_s$
in the both cases and coincides with that obtained in the weak absorption limit.

Strictly speaking, the assumed quasi-particle decay, that implies
infinite density of the final states in the background, seems to destroys the unitarity of the scattering matrix in contradiction with what has been stated
before. In the Efetov's limit the resulting expressions for the transition probabilities are formally identical to those obtained \cite{Fyodorov2005} in the
case of the analog 2D microwave resonators with resistive walls \cite{Kuhl2005}.
The walls absorb the electromagnetic energy thus widening the resonance lines
in exactly the same way as has been described above. However it has to be
stressed that the complete identity between the Maxwell equations in
the 2D microwave cavities and the Schr\"{o}dinger
equation for an electron in a mesoscopic billiard exists only in the case of perfectly reflecting walls. At the same time, quite different physics stays behind the absorption processes in these
two cases. While absorbed photons fully disappear in the environment, an electron preserves its individuality there to a certain extent. It can only lose, because of the many-body effects, a part of its energy but inevitably returns sooner or later in the cavity and  escapes finally via one of the leads.

In fact, a single-particle state in the environment with a very dense but, nevertheless, discrete spectrum is not a
stationary state with a given energy $\varepsilon_\mu$. Precisely, such a state
being once excited evolves after that quite similar to a quasi-stationary state
till the time $2\pi/\delta\gg \tau_e=1/\Gamma_e$. Only after this time recovery of the initial non-stationary state begins. A good probability there exists for an electron to be re-emitted in the cavity with some energy $E_{out}<E_{in}\approx E_{res}$ within the much shorter time interval $\tau_d=\frac{2\pi}{d}$. In an individual event of scattering with a given energy $E_{in}$ such an electron does not make resonant contribution if $E_{res}-E_{out}>\Gamma+\Gamma_s$ and therefore escapes observation. The portion of energy lost by such a retarded electron dissipates inside the environment. As a result, the background temperature jumps slightly up during each act of the scattering. However, supposing that the environment system is bulky enough, we can disregard the corresponding very slow increase of the environment temperature. Alternatively, we can suppose that a special cooling technique is in 
use. 

\subsection{Overlapping doorway resonances}
In the regime of overlapping doorway resonances, the fine-scale
averaged cross section reads (see Eq. (\ref{T_DwayRep}))
\begin{equation}\label{AvTrCrossSecOv}
\overline{\sigma^{ab}(E)}=
\sum_{n' n} {{\cal A}_{n'}^a}^* {{\cal A}_{n'}^b}^* {\cal A}_n^a
{\cal A}_n^b\,\overline{\frac{1}{{\cal D}^*_{n'}(E){\cal D}_n(E)}}\,.
\end{equation}
A little tedious calculation yields
\begin{equation}\label{AvDen}
\fl \overline{\frac{1}{{\cal D}^*_{n'}(E){\cal D}_n(E)}}
=\frac{1}{{\tilde{\cal D}}^*_{n'}(E){\tilde{\cal D}}_n(E)}\left[1+\frac{\Gamma_s^2}{-i\Gamma_s\left({\cal E}^*_{n'}-
{\cal E}_{n}\right)+\kappa\,{\tilde{\cal D}}^*_{n'}(E){\tilde{\cal D}}_n(E)}\right]
\end{equation}
where ${\tilde{\cal D}}_n(E)=E-E_n+
\frac{i}{2}\left(\Gamma_n+\Gamma_s\right)$

Let us consider at first the idealized case $\kappa=0$. In that case an incoming  electron with an energy $E_{in}$ excites via the doorway states a long-lived single-(quasi)electron state in the environment. Such a quasi-particle escapes from the device in two steps: at first, it repopulates a number of the doorway states in the cavity after a while, that decay, finally, through the leads. Appearance of mesoscopic fluctuations in processes of such a kind has been demonstrated in \cite{Amir2008}.

Taking into account the two following identities:
$$\begin{array}{c}
\frac{i}{{\cal E}^*_{n'}-{\cal E}_{n}}=\int_0^{\infty} dt_r
e^{i\left({\cal E}^*_{n'}-{\cal E}_{n}\right) t_r};\\
\frac{e^{-i{\cal E}_n t_r}}{{\tilde{\cal D}_n(E)}}=
-ie^{-i\left(E+\frac{i}{2}\Gamma_s\right) t_r}\int_{t_r}^\infty dt\,
e^{iEt-i\left({\cal E}_n-\frac{i}{2}\Gamma_s\right)t}\\
\end{array}$$
we represent the fine-structure-averaged cross section (\ref{AvTrCrossSecOv}) as a sum, $\overline{\sigma^{ab}(E)}=\sigma_d^{ab}(E)+\sigma_r^{ab}(E)$, of incoherent flows the first of which,
\begin{equation}\label{Sigma_d}
\sigma_d^{ab}(E)=
\Big|\sum_n\frac{{\cal A}_n^a {\cal A}_n^b}
{{\tilde{\cal D}_n(E)}}\Big|^2\,,
\end{equation}
describes the contribution of the overlapping doorway states
damped because of the electron capture by the environment when
the second one,
\begin{equation}\label{Sigma_r}
\sigma_r^{ab}(E)=\Gamma_s\int_0^\infty dt_r\sigma_r^{ab}(E;t_r),
\sigma_r^{ab}(E;t_r)=\Big|\sum_n\frac{{\cal A}_n^a {\cal A}_n^b }{{\tilde{\cal D}}_n(E)}\,e^{-i{\cal E}_n t_r}\Big|^2\,,
\end{equation}
accounts for the particles that spend some time $t_r$ in the
background, repopulate the doorway levels, and finally escape
via the channel $a$. Particles delayed for different times
$t_r$ contribute incoherently.

With the help of the Bell-Steinberger relation
$$\frac{1}{{\cal E}^*_{n'}-{\cal E}_n}=
-i\frac{U_{n'n}}{\sum_c {\cal A}_{n'}^c {\cal A}_n^c}$$
contribution $G_r(E)=\sum_{a\in 1,d\in 2}\sigma_r^{ab}(E)$ of
the re-injected  particles can be transformed to
\begin{equation}\label{TranspN}
G_r(E)=\sum_{n'n}U_{n'n}\sqrt{U_{n'n'}U_{n n}}\,
\frac{\sum_{a\in 1}{\Phi_{n'}^a}^*\,\Phi_n^a\,\,
\sum_{b\in 2}{\Phi_{n'}^b}^*\,\Phi_n^b }{\sum_{a\in 1}{\Phi_{n'}^a}^*\,
\Phi_n^a + \sum_{b\in 2}{\Phi_{n'}^b}^* \,\Phi_n^b }\,.
\end{equation}
Here $U=\Psi^{\dag}\Psi$ is the matrix of non-orthogonality of
the overlapping doorway states and the subsidiary amplitudes
\begin{equation}\label{PhiAmp}
\Phi_n^a(E)=\frac{\sqrt{\Gamma_s}\,\,
{\cal A}_n^a/\sqrt U_{n n}}{{\tilde{\cal D}}_n(E)}
\end{equation}
implicate the fictitious channel. The quantities $\Gamma^a=
\frac{1}{U_{n n}}|{\cal A}_n^a|^2$ satisfy the condition
$\sum_a\Gamma^a=\Gamma$ and are the partial
widths. The returning particles cannot escape directly but
rather repopulate before the doorway states that finally decay
through the external channels. In the case of moderately
overlapping doorway resonances the matrix $U_{n' n}\approx
\delta_{n' n}$ and only the terms that contain the
probabilities $\big|\Phi_n^a(E)\big|^2$ contribute. The result
obtained in such a way is a direct extension of Eq.
(\ref{Transp1}). However, contributions of different doorway
resonances in (\ref{TranspN}) nevertheless interfere when the overlap is
strong.

\section{Ensemble averaging}
Since the electron motion in the cavity is supposed to be
classically chaotic the ensemble averaging $\langle ...\rangle$
in the doorway sector is appropriate. It is easy to see that, as
long as the inelastic effects in the background are fully neglected,
such an averaging perfectly eliminates dependance of all mean cross
sections on the spreading width. Indeed, the ensemble averaged cross
section (\ref{Sigma_d}) is expressed in the terms of the S-matrix
two-point correlation function $C_V^{a b}(\varepsilon)=C_0^{a
b}(\varepsilon-i\Gamma_s)$ as
\begin{equation}\label{AnsEvCrossSec_d}
\langle\sigma_d^{ab}(E)\rangle=C_V^{a b}(0)=C_0^{a b}(-i\Gamma_s)=
\int_0^{\infty} dt\,e^{-\Gamma_s t}\,K_0^{a b}(t)\,.
\end{equation}
The subscript $V$ indicates the coupling to the background and
the function $K_0^{a b}(t)$ is the Fourier transform of the
correlation function $C_0^{a b}(\varepsilon)$. On the other
hand, using the identity
$$\frac{e^{-i{\cal E}_n t_r}}{{\tilde{\cal D}_n(E)}}=\frac{1}{2\pi}\,
e^{\Gamma_s t_r/2}\int_0^{\infty} dt\,e^{iEt}\,
\int_{-\infty}^{\infty} dE'\,
\frac{e^{-i E' (t+t_r)}}{{\tilde{\cal D}_n(E')}}$$
one can convince oneself that
\begin{equation}\label{AnsEvCrossSec_r}
\langle\sigma_r^{ab}(E;t_r)\rangle=\int_0^{\infty} dt\,
e^{-\Gamma_s t}\,K_0^{a b}(t+t_r)\,.
\end{equation}
Therefore, finally,
\begin{equation}\label{AnsEvCrossSec}
\begin{array}{c}
\langle\overline{\sigma^{ab}(E)}\rangle=
\int_0^{\infty} dt\,e^{-\Gamma_s t}\,K_0^{a b}(t)\\+
\Gamma_s\int_0^{\infty} dt_r\int_0^{\infty} dt\,e^{-\Gamma_s t}\,
K_0^{a b}(t+t_r)=
\int_0^{\infty} dt\,\,K_0^{a b}(t)=\langle\sigma_0^{ab}(E)\rangle \,.
\end{array}
\end{equation}
The ensemble averaging, being equivalent to the energy
averaging over the doorway scale $D$, suppresses all
interference effects save the elastic enhancement because of
the time reversal symmetry, which manifests itself in
the weak localization phenomenon. This symmetry is violated only
owing to the energy absorption in the environment.

\section{Quasi-particle decay and suppression of the weak localization}
Reverting now to Eqs. (\ref{AvTrCrossSecOv}, \ref{AvDen}) we
rewrite the ensemble-averaged cross sections as
$\langle\overline{\sigma^{ab}(E)}\rangle=\sigma_0^{ab}(E)+
\Delta\sigma^{ab}(E;\kappa)$ where the correction caused by the
absorption looks as
\begin{equation}\label{AbsCorr}
\Delta\sigma^{ab}(E;\kappa)=
\kappa\sum_{n' n}\langle\frac{{{\cal A}_{n'}^a}^* {{\cal A}_{n'}^b}^*
{\cal A}_n^a {\cal A}_n^b}{{\cal E}^*_{n'}-
{\cal E}_{n}}\,
\frac{1}{{\cal E}^*_{n'}-
{\cal E}_{n}+i\frac{\kappa}{\Gamma_s}
\tilde{\cal D}^*_{n'}(E){\tilde{\cal D}}_n(E)}\rangle\,.
\end{equation}
It turns out that this correction (that is not at all
necessarily positive definite) can still be expressed in terms
of the Fourier transform $K_0^{a b}(t)$ of the two-point
correlation function $C_0^{a b}(\varepsilon)$. To show this we
expand at first the expression (\ref{AbsCorr}) into power series
with respect to the parameter $\kappa$ and then make use of the
relations:
$$\begin{array}{c}
\frac{1}{\left({\cal E}^*_{n'}-
{\cal E}_{n}\right)^{(k+1)}}=-\frac{i}{k!}\int_0^{\infty} dt\,
e^{\Gamma_s t}\,(-it)^k\,
e^{-i\tilde{\cal D}^*_{n'}(E) t}\,e^{i\tilde{\cal D}_{n}(E) t};\\
\tilde{\cal D}_{n}^k(E)\,e^{i\tilde{\cal D}_{n}(E) t}=
\left(-i\frac{d}
{d t}\right)^k\, e^{i\tilde{\cal D}_{n}(E) t};\\
e^{i\tilde{\cal D}_{n}(E) t}=-e^{i E t}\frac{1}{2\pi i}
\int_{-\infty}^{\infty} dE'\frac{e^{-i E' t}}{\tilde{\cal D}_{n}(E')}\,.
\end{array}$$
Subsequent summation brings us to the result
$$\begin{array}{c}
\Delta\sigma^{ab}(E;\kappa)=
-\frac{\kappa}{(2\pi)^2}
\int_0^{\infty} dt_r\int_0^{\infty} dt\,e^{\Gamma_s(t_r+t)}
\left[e^{
-\frac{\kappa t}{\Gamma_s}\frac{\partial^2}{\partial t_1\partial t_2}}
\,e^{i E(t_1-t_2)}\right.\\
\times\left.\int_0^{\infty} dE_1\int_0^{\infty} dE_2\,
e^{i E_1(t_r+t_2)-E_2(t_r+t_1)}
C_0\left(E_1-E_2-i\Gamma_s\right)\right]_{t_1=t_2=t}\,.\\
\end{array}$$

After a change of variables:
$$
\begin{array}{cc}
\overline{E}=\frac{1}{2}(E_1+E_2),\quad & \quad \overline{t}=
\frac{1}{2}(t_1+t_2), \\
\varepsilon=E_1-E_2 \quad & \quad \tau=t_1-t_2\\
\end{array}$$
and integration over the variables $\overline{E}$ and
$\varepsilon$ this yields
$$\begin{array}{c}
\Delta\sigma^{ab}(E;\kappa)=-\sqrt{\frac{\kappa \Gamma_s}{4\pi}}
\int_0^{\infty} dt_r\int_0^{\infty}\frac{dt}{\sqrt{t}}\,
e^{-t\left[-\frac{d}{dt_r}+\frac{\kappa}{4\Gamma_s}
\left(\frac{d}{dt_r}-\Gamma_s\right)^2\right]}
K_0^{a b}(t_r)\\
=-\sqrt{\frac{\kappa \Gamma_s}{4}}\int_0^{\infty}
\frac{dt_r}{\sqrt{-\frac{d}{dt_r}+\frac{\kappa}{4\Gamma_s}
\left(\frac{d}{dt_r}-\Gamma_s\right)^2}}\,K_0^{a b}(t_r)\,.\\
\end{array}$$
(Notice that the form-factors $K_0^{a b}(t)$ monotonously
decrease with the time $t$.) Being presented in such a form
this expression is equally valid for both the orthogonal (GOE)
as well as the unitary (GUE) symmetry classes.

To simplify subsequent calculation we will consider the case of
an appreciably large number $M\gg 1$ of statistically
equivalent scattering channels, all of them with the maximal
transmission coefficient $T=1$. Then the channel indices $a, b$
can be dropped. The characteristic decay time
$t_W=1/\Gamma_W=\tau_D/M$ (dwell time) of the function $K(t)$
is much shorter than the mean delay time $\tau_D$ (here
$\Gamma_W=\frac{D}{2\pi}M$ is the so called Weisskopf width).
It is convenient to represent the function $K_0(t)$ that is
real, positive definite, monotonously decreases with the time
$t$ and satisfies the conditions $K_0(t<0)=0,\,\,K_0(0)=1$ in
the form of the mean-weighted decay exponent \cite{Sokolov2007}
\begin{equation}\label{LaplRepr}
K_0(t)=\int_0^{\infty}d\Gamma\,
e^{-\Gamma t}\,w(\Gamma),\quad \int_0^{\infty}d\Gamma\,w(\Gamma)=
K_0(0)=1\,.
\end{equation}
Rigorously, the weight functions $w(\Gamma)$ have different
forms before ($t<\tau_D$) and after ($t>\tau_D$)
the mean delay (Heisenberg) time $\tau_D$. However contribution
of the latter interval is small as $e^{-M}$ \cite{Sokolov2007}.
Neglecting such a contribution we obtain in any inelastic channel
\begin{equation}\label{Final_2}
\Delta\sigma(E;\kappa)=
-\sqrt{\frac{\kappa \Gamma_s}{4}}
\int_0^{\infty}
d\Gamma\,\frac{w(\Gamma)}{\Gamma}\frac{1}{\sqrt{\Gamma+
\frac{\kappa}{4\Gamma_s}(\Gamma+\Gamma_s)^2}}\,.
\end{equation}
In the strong absorption limit
$\kappa\gg\frac{4\Gamma_s\Gamma_W}{(\Gamma_s+\Gamma_W)^2}$ the
parameter $\kappa$ disappears from the found expression and the latter reduces to
\begin{equation}\label{InfAbsCor}
\fl \Delta\sigma(E;\kappa)\Rightarrow -\Gamma_s\int_0^{\infty}
d\Gamma\,\frac{w(\Gamma)}{\Gamma\,(\Gamma+\Gamma_s)}
=-\Gamma_s\int_0^{\infty} dt_r\int_0^{\infty} dt\,e^{-\Gamma_s t}\,
K_0^{a b}(t+t_r)\,.
\end{equation}
According to Eqs. (\ref{AnsEvCrossSec_d}, \ref{AnsEvCrossSec}) this brings us
to the result
\begin{equation}\label{Efet's_Limit}
\langle\overline{\sigma(E)}\rangle=\int_0^{\infty} dt\,
e^{-\Gamma_s t}\,K_0(t)=\int_0^{\infty}d\Gamma\frac{w(\Gamma)}{\Gamma+\Gamma_s}=
\langle\sigma_d(E)\rangle
\end{equation}
identical to this of the Efetov's imaginary-potential model \cite{Efetov1995}.
It follows from (\ref{Efet's_Limit}) that the averaged cross section approaches $1/\Gamma_s$ independently of the symmetry class if the spreading width $\Gamma_s$ noticeably exceeds the typical widths contributing to the integral over $\Gamma$. 
 
\begin{figure}
\includegraphics
{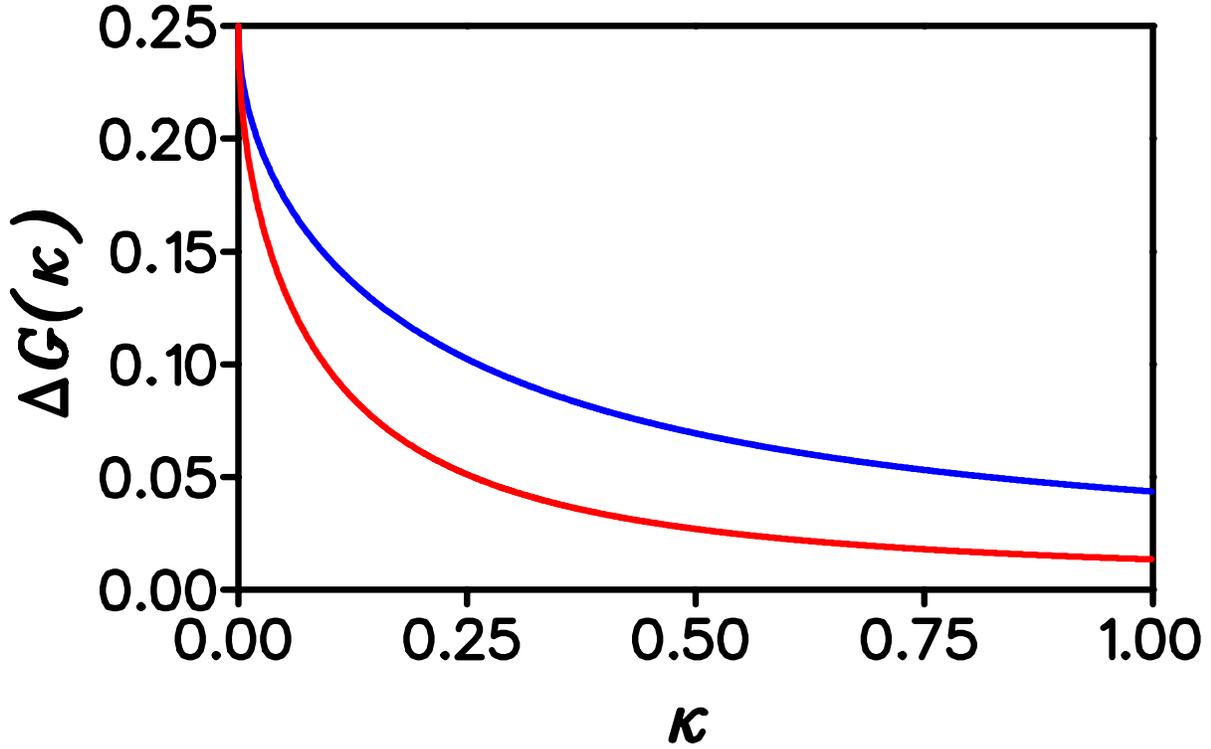}
\caption{Weak localization versus absorption parameter $\kappa$. Blue line
$\gamma_s=25$, red line $\gamma_s=64$. } 
\label{fig:WeakLoc}
\end{figure}

In the opposite limit of weak absorbtion
$\kappa\ll\frac{4\Gamma_s\Gamma_W}{(\Gamma_s+\Gamma_W)^2}$ the
expression (\ref{Final_2}) reads
\begin{equation}\label{WeakAbsCor}
\Delta\sigma(E;\kappa)\Rightarrow -\sqrt{\frac{\kappa\Gamma_s}{4}}
\int_0^{\infty}
d\Gamma\,\frac{w(\Gamma)}{\Gamma^{\frac{3}{2}}}
\end{equation}
so that
\begin{equation}\label{WeakAbs}
\langle\overline{\sigma(E)}\rangle=\int_0^{\infty}d\Gamma\frac{w(\Gamma)}{\Gamma}
\left(1-\sqrt{\frac{\kappa\Gamma_s}{4\Gamma}}\right)\,.
\end{equation}
In the case of time reversal symmetry (GOE) the asymptotic expansion
\cite{Verbaarschot1985} of the two-point correlation function gives \cite{Sokolov2007}
\begin{equation}\label{W_TRS}
w^{(GOE)}(\Gamma)=\delta(\Gamma-\Gamma_W)-
\frac{2}{t_H}\delta'(\Gamma-\Gamma_W)
+\frac{M}{2t_H^2}\delta''(\Gamma-\Gamma_W)+...
\end{equation}
whereas in the case of absence of such a symmetry (GUE)
\begin{equation}\label{W_no_TRS}
w^{(GUE)}(\Gamma)=\delta(\Gamma-\Gamma_W)+...\,.
\end{equation}
In the both cases contributions of the omitted terms are
$O(1/(M^{-7/2}))$. With such an accuracy, the formula (\ref{Final_2})
yields for the weak localization the expression
\begin{equation}\label{weak_loc_gen}
\begin{array}{c}
\Delta G\equiv G^{(GUE)}-G^{(GOE)}\\
= M_1M_2\left(2\frac{d}{d\mu}+
\frac{\mu}{2}\frac{d^2}{d\mu^2}\right)
\left\{\frac{1}{\mu}\left[1-\frac{\sqrt{\frac{\kappa\gamma_s}{4}}}
{\sqrt{\mu+\frac{\kappa}{4\gamma_s}(\mu+\gamma_s)^2}}\right]\right\}\Big|_{\mu=M}\\
\end{array}
\end{equation}  
which is valid for arbitrary values of the parameters $\kappa$, $\gamma_s$ 
and $M$. The unfolded explicit expression is a bit too lengthy. Therefore we visualize this result in Fig.~\ref{fig:WeakLoc} for two different values of the (dimensionless) spreading width $\gamma_s$ and $M_1=M_2=2; M=4$. 

In fact, the condition $\gamma_s\gtrsim M$ can hardly hold if the number of
channels is very large, $M\gg 1$. In this case a natural estimate of the absorption
parameter is $\kappa\approx 4\frac{\Gamma_s}{\Gamma_W}=4\frac{\gamma_s}{M}$. With this accuracy we obtain for the transport probabilities:
\begin{equation}\label{Transp2}
\begin{array}{c}
G^{(GOE)}=\frac{M_1M_2}{M}\left[\left(1-
\frac{\gamma_s}{M}\right)-\frac{1}{M}\left(1-\frac{9}{8}
\frac{\gamma_s}{M}\right)\right],\\
G^{(GUE)}=\frac{M_1M_2}{M}\left(1-
\frac{\gamma_s}{M}\right).\\
\end{array}
\end{equation}
The principal terms are identical and the difference
\begin{equation}\label{eq12}
\Delta G\equiv G^{(GUE)}-G^{(GOE)}=
\frac{M_1M_2}{M^2}\left(1-
\frac{9}{8}\frac{\gamma_s}{M}\right)
\end{equation}
describes a slight suppression of the weak localization.

The effect of decoherence becomes more and more pronounced as the number of channels
decreases. However the asymptotic expansions (\ref{W_TRS}, \ref{W_no_TRS})
are not justified in the case of few number of channels. Besides, violation
of analyticity of the function $K_0(t)$ at the point $t=\tau_D$ becomes
essential \cite{Sokolov2007}. The problem calls for a special consideration
in this case.

\section{Summary}
In this paper we have described a possible mechanism of
decoherence and absorption phenomena in the electron quantum
transport through an open ballistic 2D mesoscopic cavity. These effects
are induced by a weak time-independent interaction with a bulky
disordered environment with a very dense but, nevertheless,
discrete spectrum. Due to such an interaction, each doorway resonance
state in the cavity gets fragmented onto a large number
$\sim\Gamma_s/d$ (the spreading width $\Gamma_s$ characterizes the
strength of the coupling to the environment when $d$ is the
single-quasi-particle mean level spacing) of very narrow resonances
that cannot be resolved experimentally. Only the cross sections averaged
over the fine $(\thicksim d)$ structure scale are measured. The observable
cross sections at a given energy of incoming electron $E_{in}$ turn out to be incoherent sums of flows the first of which corresponds to the scattering with excitation and subsequent decay of the doorway resonances broadened
because of the internal friction induced by interaction with the
environment. Such a broadening by the spreading width $\Gamma_s$ simulates absorption. The second flow accounts for the particles re-injected in the
cavity after some retardation time spent in the environment. The degree of
decoherence of the electrons contributing to these two flows is described 
by the decoherence rate $\gamma_{\phi}=\gamma_s=\Gamma_s\tau_D$ uniquely defined 
by the spreading width.

Formally, the transition amplitudes (\ref{T_DwayRep}) obey the unitarity
condition at any given scattering energy $E$. This implies, in particular,
that the number of electrons is conserved during the stationary scattering.
All of them enter into and escape from the cavity only through the attached open leads. However long-lasting $(\tau\thicksim\tau_{\delta}=\frac{2\pi}{\delta}$) evolution of the unsteady state of a (quasi)electron once penetrated into the many body environment entails loss of the electron's energy in any individual act of scattering lasting a much shorter time $\sim\tau_d=\frac{2\pi}{d}\lll \tau_{\delta}$. So, some electrons leave the cavity with energies well below the energy of the incoming particles and elude observation made at the resonant scattering energy $E_{res}\approx E_{in}$ (fixed with accuracy worse than $d$). Rigorously speaking, the energy dissipation causes slow heating of the environment. We disregard the latter effect supposing that the environment is bulky enough or, alternatively, a special cooling procedure is used. Disappearance of the particles from the resonant scattering interval not only imitates a seeming loss of particles but yields also real violation of the time reversal symmetry and, as a consequence, suppression of the weak localization the both effects being controlled by one and only  parameter - the decoherence rate $\gamma_s$.

\ack
I am very obliged to D.V. Savin for his interest to this work,
numerous discussions and advice. Also, I am very grateful to
C.W.J. Beenakker and P.A. Mello for discussions and criticism.
Financial support by RFBR (grant ¹ 09-02-01443) and by the RAS
Joint scientific program "Nonlinear dynamics and Solitons" is
acknowledged with thanks.

\end{document}